\documentclass[aps,pra,showpacs,twocolumn,superscriptaddress]{revtex4}
\usepackage{bm}
\usepackage[usenames]{color}
\usepackage{epsfig}
\usepackage{times}
\usepackage{amssymb}

\begin{document}
%%%%%%%%%%%%%%%%%%%%%%%%%%%%%%%%%%%%%%%%%%%%%%%%%

\bigskip

\bigskip

\title{Quantum and Classical Calculations of Ground State Properties of \\
Parabolic Quantum Dots. }

\author{V.~Popsueva}
\affiliation{Department of Physics and Technology, University
  of Bergen, N-5007 Bergen, Norway}

\affiliation{Center of Mathematics for Applications, University of Oslo,  N-0316
Oslo, Norway}

\author{T.~Matthey}
\affiliation{Bergen Center for Computational Science,
N-5008 Bergen,  Norway}

\author{J.~P.~Hansen}
\affiliation{Department of Physics and Technology, University
  of Bergen, N-5007 Bergen, Norway}

\author{L.~Kocbach}
\affiliation{Department of Physics and Technology, University
  of Bergen, N-5007 Bergen, Norway}

\author{M.~Hjorth-Jensen}
\affiliation{Department of Physics and Center of Mathematics for Applications,
University of Oslo,  N-0316 Oslo, Norway}
\affiliation{Department of Physics and Astronomy, Michigan State
University, East Lansing, Michigan 48824, USA}

% Abstract

\begin{abstract}

We report calculations for electronic ground states of parabolically
confined quantum dots for up to 30 electrons  based on the quantum
Monte Carlo method. Effects of the electron-electron interaction and
the response to a magnetic field are exposed. The wavefunctions and
the ground state energies are compared with purely classical
calculations performed with a comprehensive Molecular Dynamics code.
For the chosen well parameters a close correspondence in the overall
shape of electron density distribution is found even for small
number of electrons, while the detailed radial distributions show
the effects of Pauli principle in the quantal case.

\end{abstract}

\maketitle

\section{Introduction}

% General q-dots & technology

The experimental and technological advances in production and use
of two-dimensional quantum dot structures
provide also new challenges and possibilities
for theoretical studies reaching into surprisingly wide areas
of physics
\cite{Reimann-review}. For example, techniques for monitoring
of controlled single-electron coherent transport through double
dot systems has recently been realized \cite{hayashi03}. Studies
of structural properties of few- \cite{harju02,petta04} and
many-electron systems
\cite{lieb95,harju99,harju05, guclu03} are in this perspective of crucial importance for
experimental progress. In particular coherent and quasi-coherent
dynamics of quantum dots with a large number of electrons pose both
theoretical and experimental challenges. 'Large' is here used in
the sense 'larger than two', but still too few to justify
application of mean field approaches \cite{book1}. Coherent
dynamics and the interaction with a classical environment leading
to  decoherence  are so far a highly unexplored area of the border
between quantum and classical physics.

% General on q-dot structure

Shell-filling effects in quantum dots were demonstrated as oscillations
in the addition energy spectra  with increasing number of
electrons in the dot almost ten years ago \cite{tarucha96}.  A few
years earlier the ground state energies as function of an external
magnetic field were  measured \cite{Ashoori93}. These findings
sparked off an intense activity on the applications of well-known
quantum mechanical structure theories from atomic and nuclear
physics to the new area of quantum dots.  The Hartree-Fock (HF)
method, Linear Combinations of Atomic  Orbitals  (LCAO) methods,
Density Functional Theory (DFT) and the quantum Monte-Carlo (QMC)
methods are here the most frequently applied (cf. e.g. \cite{book}) working
tools which have been used and further developed in quantum dot
studies.
% General structure methods
In structure theory it is well known that LCAO combined with excact
diagonalization rapidly becomes computationally impractical with
increasing number of electrons.  HF, DFT and QMC theories are in
general much more applicable with increasing particle size. The DFT
method method has successfully reproduced experimental addition
energy spectra \cite{Reimann99}, while similar success of the HF
method has not yet been reported. The HF method as well as mean
field models do however give rise to oscillatory addition energy
spectra \cite{Rontani99} as well.
% Survey of recent relevant works
Various QMC approaches  have investigated the role of
correlations, confinement geometry and shell effects for zero
external magnetic field, see eg. \cite{harting00, sundquist03,
pedriva00, resanen03, lee01}. The onset of a magnetic field
introduces an additional parameter in the Hamiltonian and thus an
implicit dependence of the N-electron energy levels as function of
the field. It is well known that with increasing (weak) field the
various field free energy levels depend differently on the field
which leads to an intricate spectrum of (avoided) level crossings.
It was for example recently shown that the
character of ground state of a
two-electron double dot changes from a $S=0$ type to a $S=1$
and back to $ S=0$ type as the  field strength is increased \cite{harju02}. The
behavior of many-particle systems in magnetic fields has
previously been investigated for up to 13 electrons \cite{harju99,
guclu03} by QMC methods.

% This work - paper setup:

In this paper we apply a new QMC procedure \cite{oslo} to study
the ground state properties of parabolically confined
two-dimensional quantum dots. We calculate addition energy spectra
for dots with 2 to 20 electrons and we display the behavior of the
ground state for increasing magnetic field. Further more, the
energy and the probability density are compared with purely
classical electron dynamics calculations. These are based on
the employment of non-equilibrium molecular dynamics methods where a
thermostat slowly cools the system to a near-frozen state.

In the following section we briefly describe the theoretical
models and the computational methods. The results are displayed
and discussed in section III followed by concluding remarks in
section IV.

\section{Models}

Our model Hamiltonian $\hat{H}$ of $N$ identical electrons parabolically
confined in two dimensions $r_i = (x_i,y_i)$ is given by
\begin{equation}
  \hat{H}=\sum_{i=1}^N  \left[ -\frac{\hbar^2}{2m^*}\nabla_i^2+V({\bf
      r}_i) + \sum_{j<i}^N \frac{e^2}{4\pi \epsilon
      \epsilon_0|{\bf r}_{i}-{\bf r}_{j}+\alpha|} \right].
  \label{ham}
\end{equation}
Here $m^*$ is the effective electron mass and $\epsilon$ is the
dielectric constant of the background material. Realistic
experimental values  for a two dimensional structure surrounded by
a GaAs material is $m^* = 0.067 m_e$ where $m_e$ is the electron
mass and $\epsilon = 13.5$. The parameter $\alpha$ is empirically
introduced to mimic a finite confinement size in the $\hat{z}-$
direction. A non-zero $\alpha$ will generally increase the mean
distance between real confined dot electrons and thereby reduce
the role of electron-electron interactions.  The confining
harmonic potential is defined by the frequency $\omega_0$ which in
real systems typically is $\hbar \omega_0 \sim 2 - 6$ meV.

In addition the dot electrons may be exposed to a constant magnetic
field, ${\bf B}$, in the $\hat{z}-$direction which correspond to an
electromagnetic vector potential ${\bf A} = \frac{1}{2}B \left(
-{\bf e}_x + -{\bf e}_y \right)$. The effective external single
particle potential can then be brought to the form,
\begin{equation}
  V({\bf r}_i) = \frac{1}{2} m^* \omega^2 r_i^2 + \frac{w_B}{2}\hat{L}_z,
\end{equation}
with $\hat{L}_z$ being the angular momentum operator of the
$\hat{z}-$ axis. The effective confining potential is given by $
w^2 = \omega_0^2 + \frac{\omega_B^2}{4}$, where $\omega_B =
\frac{|eB|}{2m^*} $ is the Larmor frequency. In the following we
apply reduced atomic units and set  $\omega = \hbar = m^* = 1$.

\subsection{Variational Monte Carlo}

In the contruction of the single-particle basis and the trial wave
function to be used in our variational Monte Carlo simulations we
follow closely the work of Harju {\em et al.}, see for example
Ref.~\cite{harju99,harju05}.

We thus assume  that the electrons in the quantum dot move in a
two-dimensional parabolic potential, meaning in turn that the
one-body problem is similar to the classical harmonic oscillator
problem. In scaled units, our ansatz for the single-particle basis
is given by
\begin{equation}
  \label{eq:spwf}
  \psi_{n,\pm|m|}\propto (x\pm
  iy)^{|m|}L_{(n-|m|/2)}^{|m|}(r^2)\exp{-(\frac{r^2}{2})},
\end{equation}
with $n$ being the shell index, $m$ the angular momentum and
$r^2=x^2+y^2$.  Since this is also a system with no well-defined
center of mass motion, we need, in order to obtain a translationally
invariant Hamiltonian to redefine the coordinates through
\begin{equation}
  x\pm iy = (x-x_{\mathrm{cm}})\pm \imath (y-y_{\mathrm{cm}}),
\end{equation}
where $x_{\mathrm{cm}}$ and $y_{\mathrm{cm}}$ are the center of mass coordinates.
In addition to the harmonic oscillator part for the
single-particle motion we assume as discussed above that the
electrons interact via repulsive Coulomb forces.

Our ansatz for the many-body wave function consists thus of
a Slater determinant multiplied by a two-body correlation function
$f(r_{ij})$, leading to the following form for the $N$-particle trial
wave function $\Psi_N$
\begin{equation}
  \Psi_N = \mathrm{det}[\psi_1\psi_2\dots\psi_N]
  \times\prod_{i<j}^Nf(r_{ij}),
\end{equation}
where $\psi_i$ refer to the single-particle wave functions defined
in Eq.~(\ref{eq:spwf}) with the index $i$ representing the
single-particle state $i$ with quantum numbers $n,m$ and positions
$(x_i-x_{\mathrm{cm}})\pm \imath (y_i-y_{\mathrm{cm}})$. The
single-particle wave functions depend on $x$ and $y$ and the
center of mass coordinates $x_{\mathrm{cm}}$ and
$y_{\mathrm{cm}}$.
In our approach here we do
not introduce variational parameters in the single-particle wave
functions of Eq.~(\ref{eq:spwf}), see also the discussions in
Refs.~\cite{harju99}.

For the two-body correlation function we assume one of the simplest
possible Jastrow factors, namely
\begin{equation}
  f(r_{ij}) = \exp{(\frac{a_1r_{ij}}{1+a_2 r_{ij}})},
\label{jastrow}\end{equation}
with $a_1$ and $a_2$ being variational parameters. Note that these
parameters are different for fermion pairs with parallel and
anti-parallel spins, respectively. This leads in total to four
variational parameters in our calculations. Note also that the
correlation function does not depend on the center of mass motion.
The work of Alexander and Coldwell \cite{ac1997} contains an extensive list of
correlation functions for light atoms systems.

The norm of the total wave function is not needed since it is
redundant in the Metropolis sampling. For more details on the
Metropolis sampling and variational and diffusion Monte Carlo
techniques see for example Refs.~\cite{harju99,harju05,bolton97,pudliner97}.
We apply here a standard Metropolis sampling, as
discussed in for example Refs.~\cite{harju99,harju05} where the main
problem is to integrate
\begin{equation} \label{energy2.4}
  E \left [\Psi_N \right ] = \frac{\int \Psi_N^* \hat{H} \Psi_N d
    \mathbf{x}}{\int \Psi_N^* \Psi_N d \mathbf{x}}.
\end{equation}
In our discussion below, we compare the energies obtained with the
above correlated wave function with those arising from the
non-interacting case, given by the sum over single-particle energies
determined by the harmonic oscillator function. These energies are
given by the Fock-Darwin energy spectrum \cite{fockdarwin}
\begin{equation}
  \epsilon_{mn} = \hbar \omega (n_x + n_y + |m| +1) - \frac{1}{2}
  \hbar \omega m.
\end{equation}

All simulations in this work were performed with $10^7$ Monte
Carlo cycles (and $5 \times 10^5$ thermalisation steps). There are
two variational parameters in the Jastrow factor of
Eq.~(\ref{jastrow}), $a_1 =1.0$ and $a_2 =1.5$ are found to give
consistently optimized results for all considered values of N. In
Table \ref{sim_data} we show some of the simulation results for
different values of $N$, with $a_1$ and $\beta$ as described
above.
\begin{table}[htp]
\begin{center}
\small{
  \begin{tabular}{|c|c|c|c|}
    \hline
    $N$ & Electron configuration & energy & standard deviation \\
    \hline
    2 & $\uparrow \downarrow$ & 3.15853 & 0.000408482 \\
    2 & $\uparrow \uparrow$ & 3.61921 & 7.99765e-05 \\
    3 & $\uparrow \downarrow \uparrow$ & 6.65201 & 0.000489449 \\
    6 & $3 \times (\uparrow \downarrow)$ & 21.4633 & 0.000900802 \\
    12 & $6 \times (\uparrow \downarrow)$ & 68.7905 & 0.000904738 \\
    20 & $10 \times(\uparrow \downarrow)$ & 169.413 & 0.00284978 \\
    \hline
  \end{tabular}\vspace{3mm}
}
\caption{Selected simulation results, energy and standard deviation for
different electron configurations and various $N$-values.}
\label{sim_data}
\end{center}
\end{table}

\subsection{Classical calculations}

A classical mechanics-based "molecular dynamics" approach is well
suited for massive particles at sufficient high temperatures, for example
in simulations of molecular liquids. Several classical simulations
of low temperature and confined dilute matter systems have also
been successfully applied recently, see for example Ref.~\cite{ioncrystals}. We
apply the same program in the present work except that the
$\hat{z}$-direction is kept frozen.

In essence, the dynamics is described by Newton's equation of
motion through
\begin{equation}
  \label{eq:newton}
  m_i \frac{d^2}{dt^2}{\bf{r}}_i(t) = {\bf{F}}_i(t),
  \label{newton}
\end{equation}
where the Force ${\bf F}$ is derived from Eq.~(\ref{ham}) and
modified at each time step $\Delta t$ according to,
\begin{equation}
  \bf{F}_i \leftarrow {\bf F}_i - \zeta {\bf v}_i \frac{1}{N \Delta t}.
\end{equation}
The thermostat $\zeta$ slowly drives the system from a
non-equilibrium initial configuration to the final state characterized
by a  required temperature $T'$,
\begin{equation}
  \frac{d\zeta}{d t} =  \gamma (E_k(t) - \frac{3N}{2}kT').
\end{equation}
Here $E_k$ is the kinetic energy, $k$ is Boltzmann's constant and
$\gamma$ is a parameter of order unity which plays no role for the
final structures, but regulates the computational speed of
cooling/heating of the system. The temperature $T'$ is the set
equilibrium temperature of the system.

When propagating the system in time we solve Eq.~(\ref{newton})
numerically with an appropriate time step by the leap-frog method
\cite{matthey-thesis}. Time steps and cooling parameters are
carefully adjusted to avoid the system to be locked into any local
energy minima close to the absolute minimum as described in
\cite{classic93}.

\section{Results and Discussion}

\subsection{Energy levels}
\begin{figure}[htb]
\begin{center}
\epsfxsize=8.0cm \epsfbox{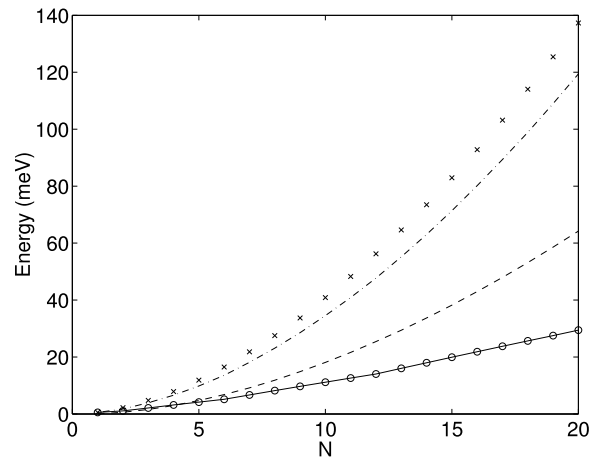}
\end{center}
\caption{\label{fig3} Ground state energy as a function of number
of electrons ($\alpha = 0, B=0$). Crosses: QMC calculation without
Jastrow factor; dashed-dotted line: QMC with the Jastrow factor.
Full line with circles: Independent particle energies. Dashed
line: Classical calculation.}
\end{figure}

In Fig.~\ref{fig3} we show the resulting ground state energies as
functions of the number of electrons $N$  and with zero magnetic
field and confinement energy $\hbar \omega_0 = 3$ meV. The two
upper curves are the variational Monte Carlo energies obtained
without and with the Jastrow factor described in the previous
section. This factor is seen to play an increasing role for
increasing number of electrons, as expected since the average
distance between the electrons decreases with increasing electrons
in the dot. For $N=20$ electrons the Jastrow factors decreases the
energy with about 10\%. To compare with the total role of the
electron-electron interaction we also plot in Fig.~\ref{fig3} the
corresponding ground state for non-interacting particles, full
line with circles. The role of the correlations is again seen to
increase with the number of electrons, but much less than for the
interacting case. Finally, in the same figure we display the
ground state energy obtained with the classical molecular dynamics
calculations. As expected the energies are a factor of two below
the quantal interacting energies due to vanishing kinetic energy
for $T=0$.

\begin{figure}[htb]
\begin{center}
\epsfxsize=8.0cm \epsfbox{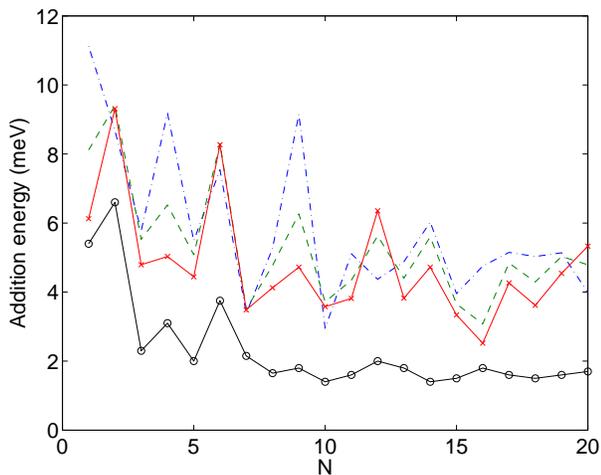}
\end{center}
\caption{\label{fig1B} (Color online) Addition energy spectrum,
computed for the
  $\hbar \omega_0 = 3 \mbox{meV}$-case, with different offsets $\alpha$. The
  blue line corresponds to $\alpha = 0$ nm. Points on the green line are
  computed with $\alpha=5$ nm, and those marked with $x$'s are with
  $\alpha=10$ nm. The points on the black line are those obtained from
  the experiment of Ref.~\cite{tarucha96}.}
\end{figure}
All calculations in Fig.~\ref{fig3} are performed with $\alpha=0$,
see Eq.~(\ref{ham}). In Fig.~\ref{fig1B} the role of screening is
examined by computing the addition energy spectra with $\hbar
\omega_0 = 3$ meV. The addition spectra is given by, $\delta \mu =
E(N+1) - 2E(N) + E(N-1)$, where $\mu$ is the chemical potential.
The energies  have been evaluated with $\alpha=0, 5$ and 10 nm.
Strong peaks at $N=2$ and $N=6$ and higher "magic" numbers are
obtained in all cases, in qualitative agreement with experiment.
The strong peak for $N=4$ gets reduced for increasing $\alpha$ and
the calculations for $\alpha = 10$nm are in better
agreement with experiment than the  $\alpha = 0$ case. One can argue for a finite
$\alpha$ value since a
quantum dot has a finite thickness, resulting in a screened Coulomb interaction between the
electrons. This leads to a reduction in the total energy and makes the
harmonic oscillator potential more dominant.
We also notice that the addition energy is large at the
half-filled shells $N=4,9,12..$. According to Hund's rule, these
are the states with maximum total spin, i.e. the electron
eigenspins are parallel. Such configurations require an
antisymmetric spatial wave function, which minimises the Coulomb
repulsion between the electrons by keeping electrons apart and
making the configuration more stable. By screeening the electrons
we reduce the effect of the Coulomb repulsion, therefore also
making the spatial antisymmetrisation less important for lowering
the energy of the configuration. Therefore, the addition energy
peaks at half-filled shells become smaller as we increase the
screening parameter $\alpha$.

\begin{figure}[htb]
\begin{center}
\epsfxsize=8.0cm \epsfbox{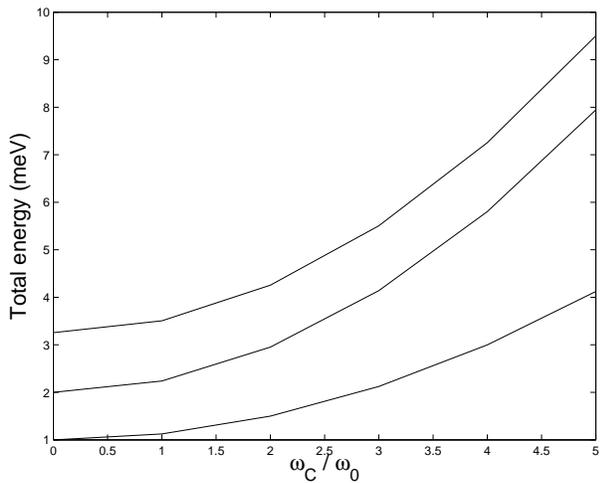}
\end{center}
\caption{\label{fig4} Total energy of a two-particle ground state
as a function of the strength of the magnetic field. Upper curve:
Two particles with the electron-electron interaction. Middle
curve: Two particles with no electron-electron interaction. Lower
curve: A single electron. The screening parameter is $\alpha =
10$.}
\end{figure}

We now turn to the non-zero magnetic field configurations. The
detailed lifting of the ground state energies with magnetic fields
is of great importance for coherent transport in coupled double
dots.  The tunneling amplitude between dots depends exponentially
on the barrier parameters and energy levels which again can be
adiabatically tuned by a time dependent external field. In
Fig.~\ref{fig4} the total energy of a two- and one-particle ground
state as a function of magnetic field strength is shown, in the
two-electron case with and without the electron-electron
interaction.  As expected, with zero total orbital momentum, the
energy increases monotonous with the magnetic field. The
contribution from the Coulomb repulsion is seen to be very
significant already for two electrons as it accounts for around
$40\%$ of the total energy.

\begin{figure}[htb]
\begin{center}
\epsfxsize=8.0cm \epsfbox{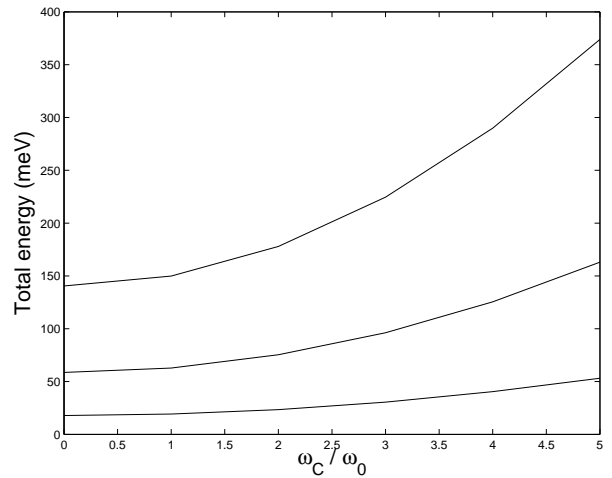}
\end{center}
\caption{\label{fig5} Total energy per particle as a function of
    magnetic field strength for $N=6,12$ and $20$. The screening
    parameter is $\alpha = 10$.}
\end{figure}

The importance of the Coulomb repulsion increases with increasing
magnetic field and number of particles. Introducing a finite
offset $\alpha$ puts a limit on how close the electrons may
approach each other. This will of course prevent the Coulomb
interaction from growing with magnetic field when the offset
$\alpha$ is large enough. In Fig.~\ref{fig5}  we plot the scaled
total energy as a function of magnetic field strength for $N=6$,
$N=12$ and $N=20$ electrons. The comparison shows that the
electron repulsion amplifies the increase of energies due to the
increasing magnetic field strength. The increased energy is seen
to be close to proportional with the number of particles.

\subsection{Electron probability distribution}

\begin{figure}
\begin{center}
\begin{tabular}{l l}
$N = 2:$ & \\
\epsfxsize=4cm \leavevmode \epsfbox{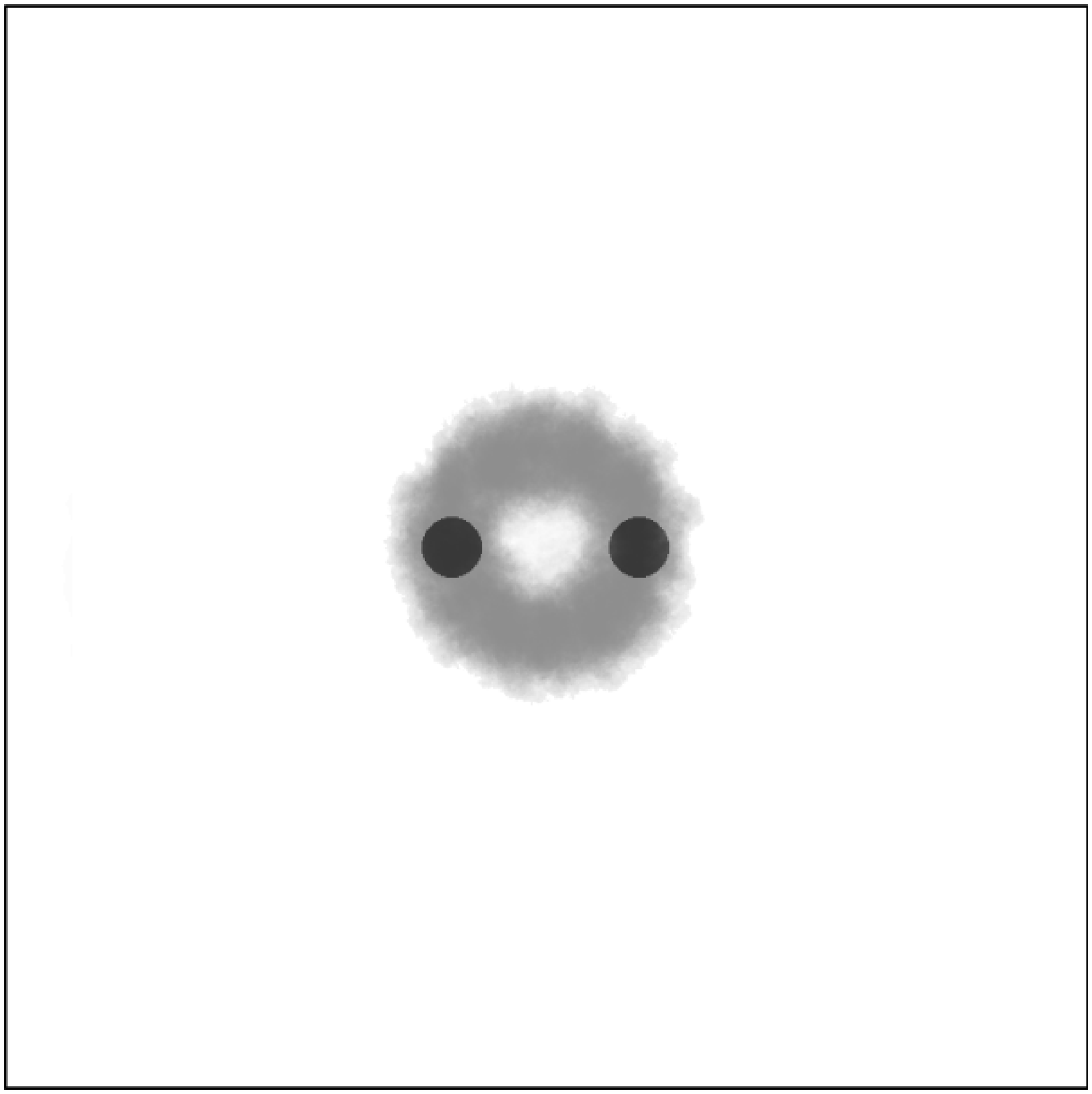} & \epsfxsize=4cm
\leavevmode
\epsfbox{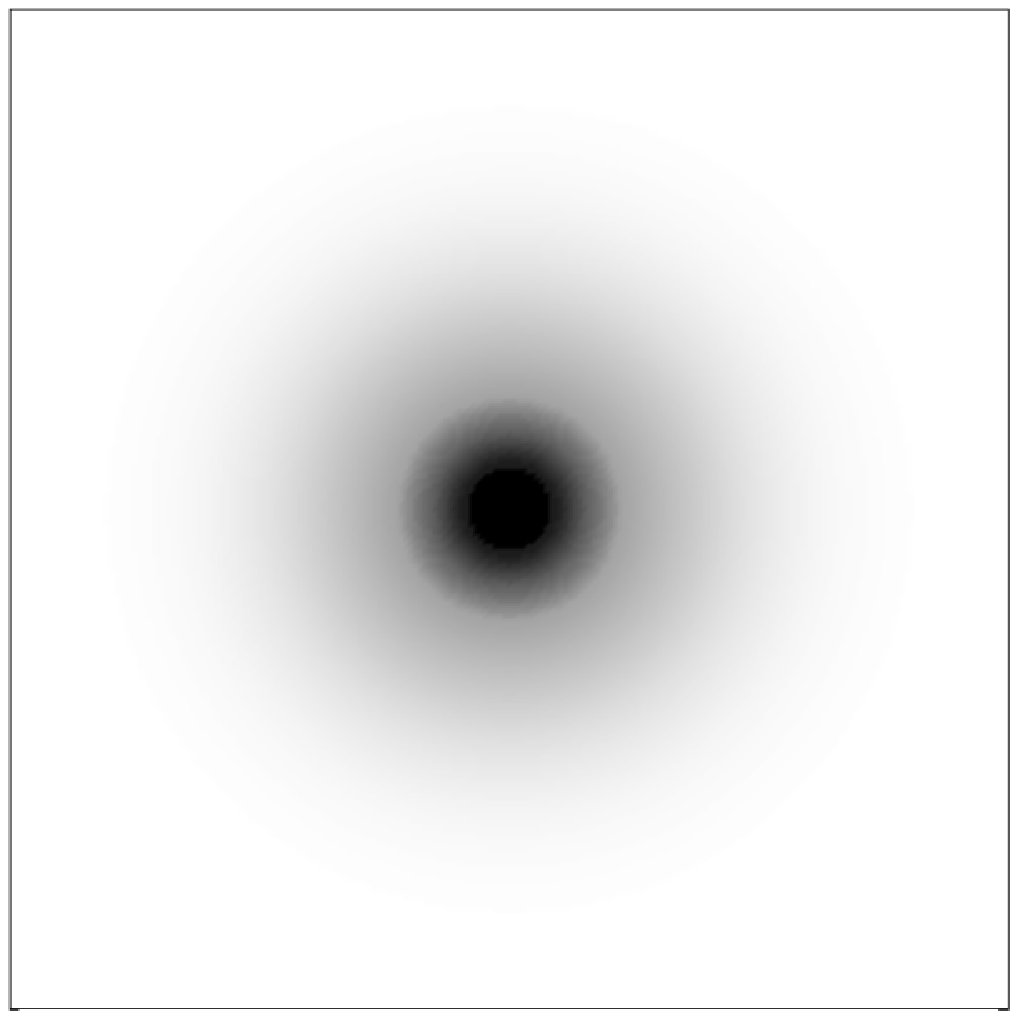}\\
$N = 6:$ & \\
\epsfxsize=4cm \leavevmode \epsfbox{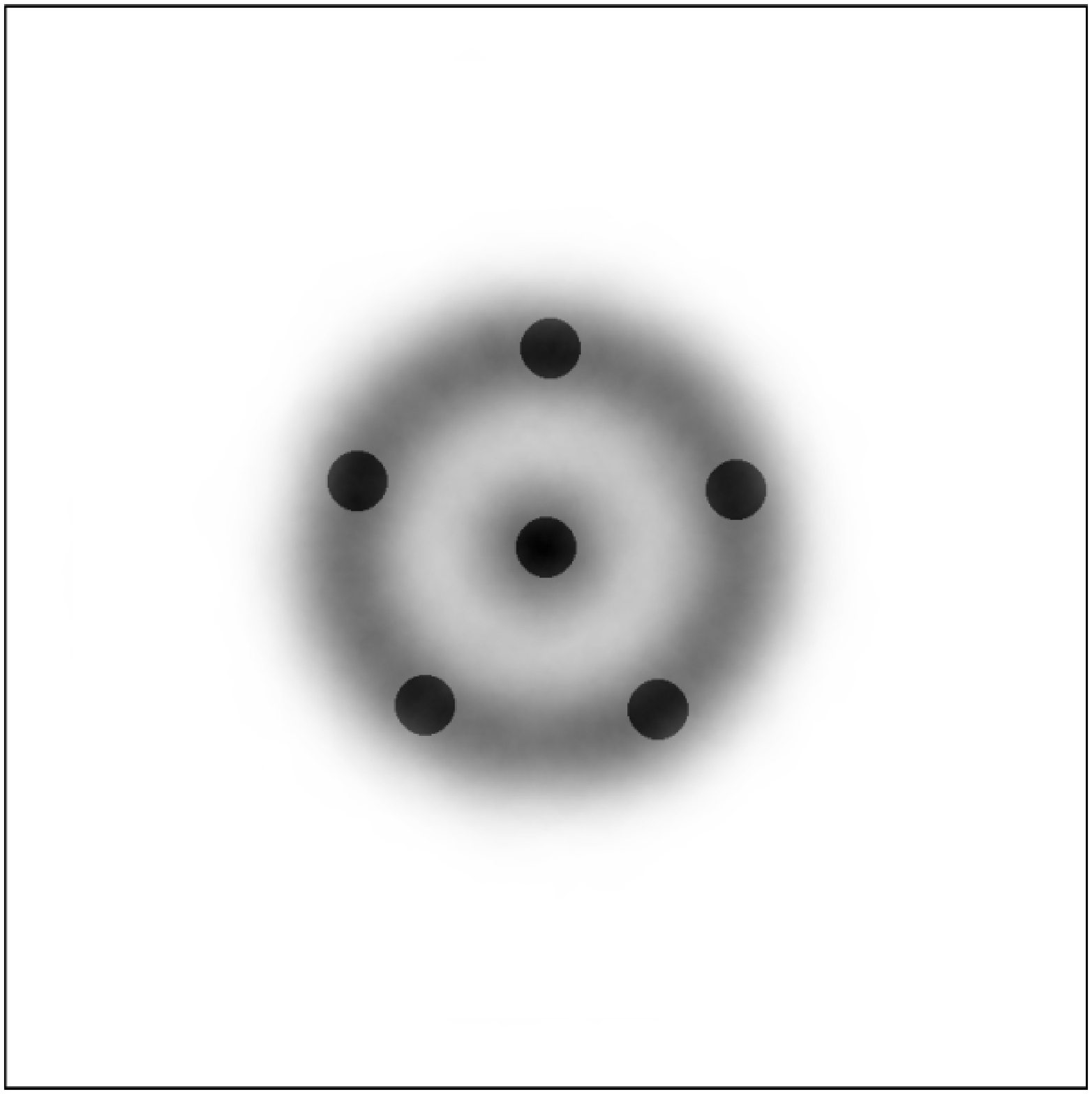} & \epsfxsize=4cm
\leavevmode
\epsfbox{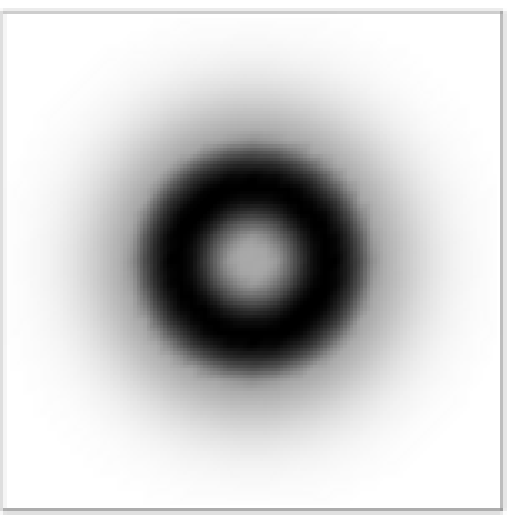}\\
$N = 12:$ & \\
\epsfxsize=4cm \leavevmode \epsfbox{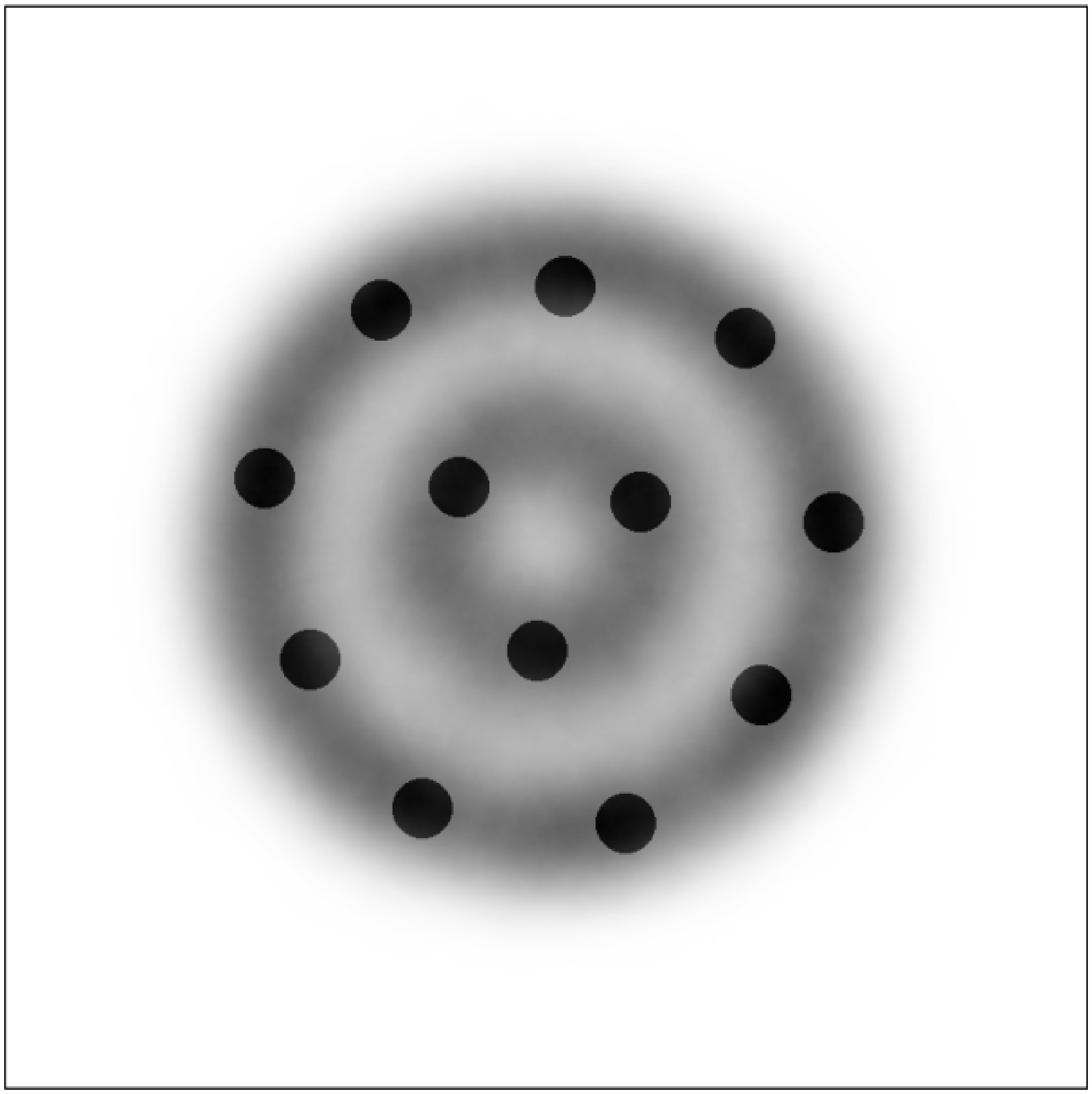} & \epsfxsize=4cm
\leavevmode
\epsfbox{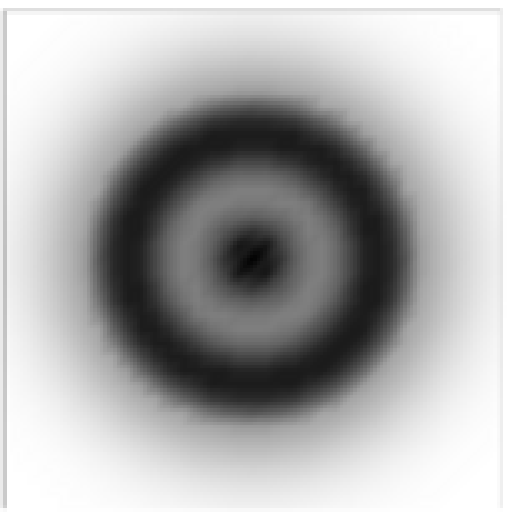}\\
$N = 20:$ & \\
\epsfxsize=4cm \leavevmode \epsfbox{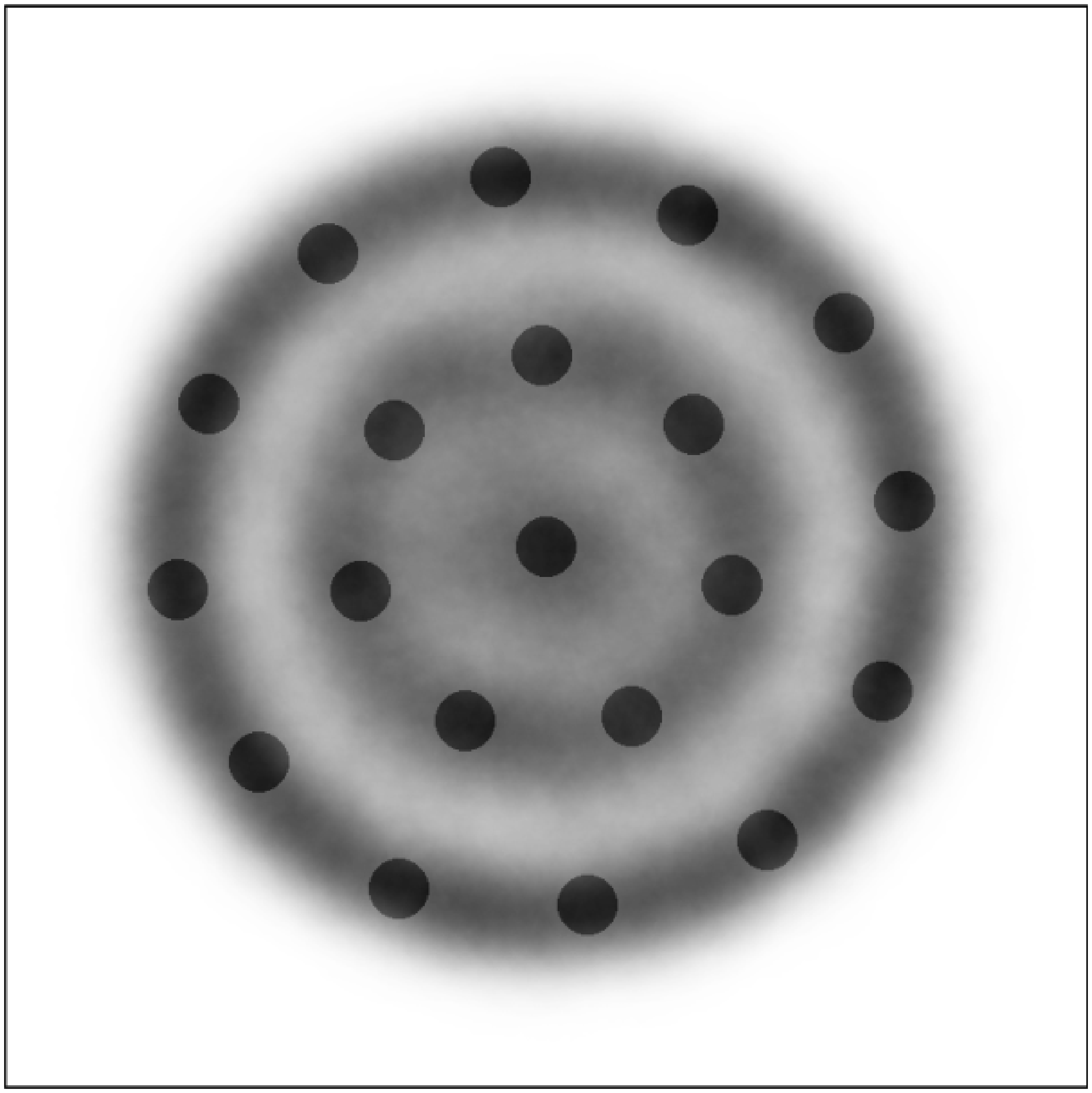} & \epsfxsize=4cm
\leavevmode
\epsfbox{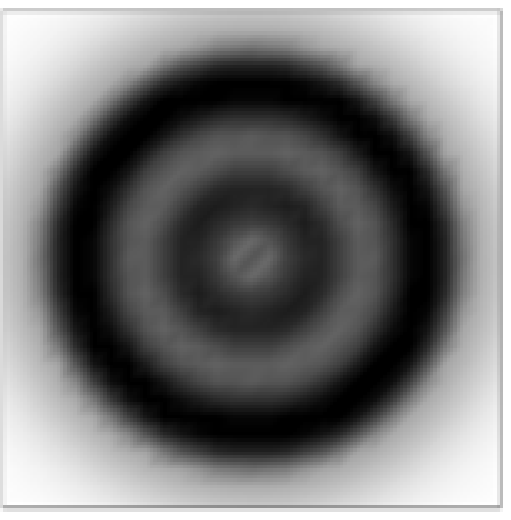}\\
$N = 30:$ & \\
\epsfxsize=4cm \leavevmode \epsfbox{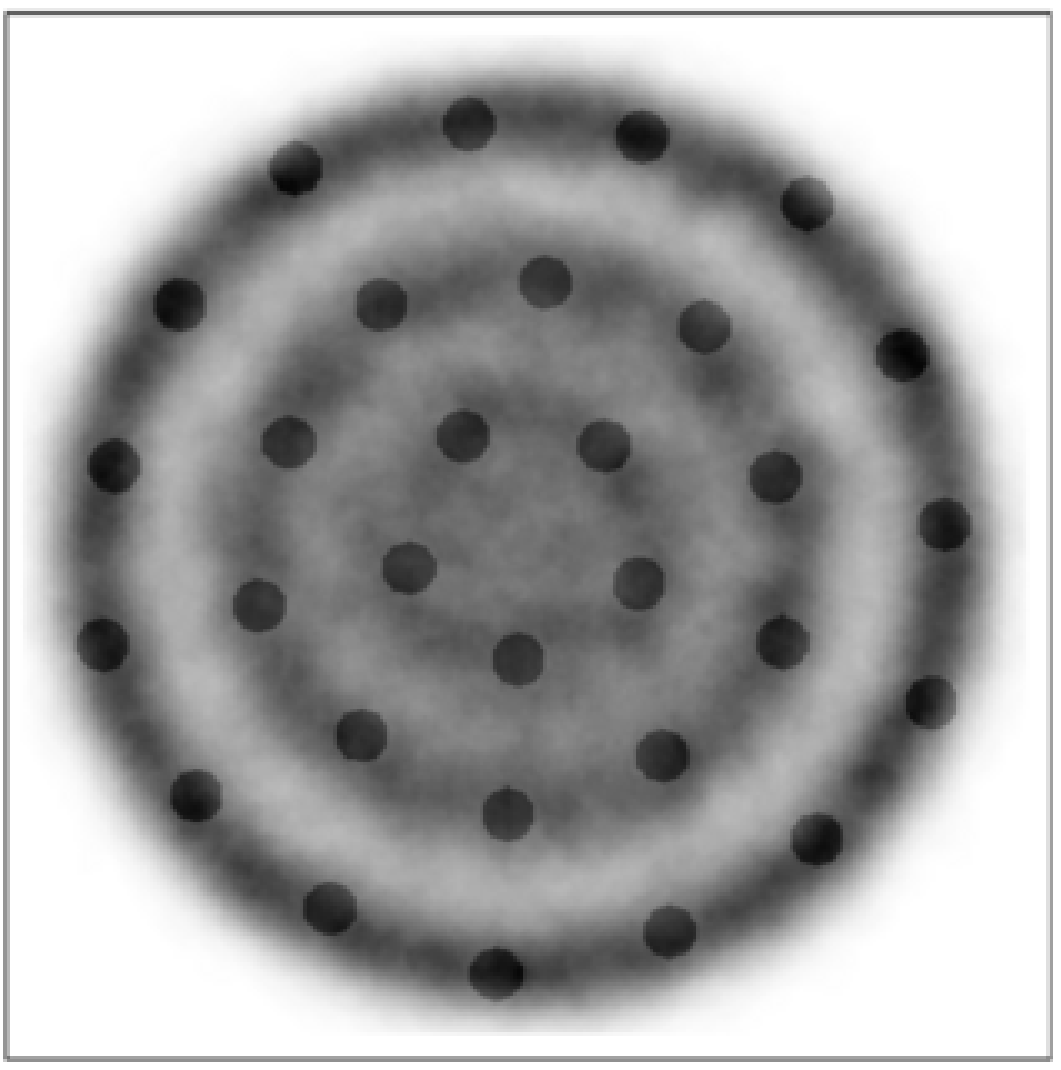} & \epsfxsize=4cm
\leavevmode
\epsfbox{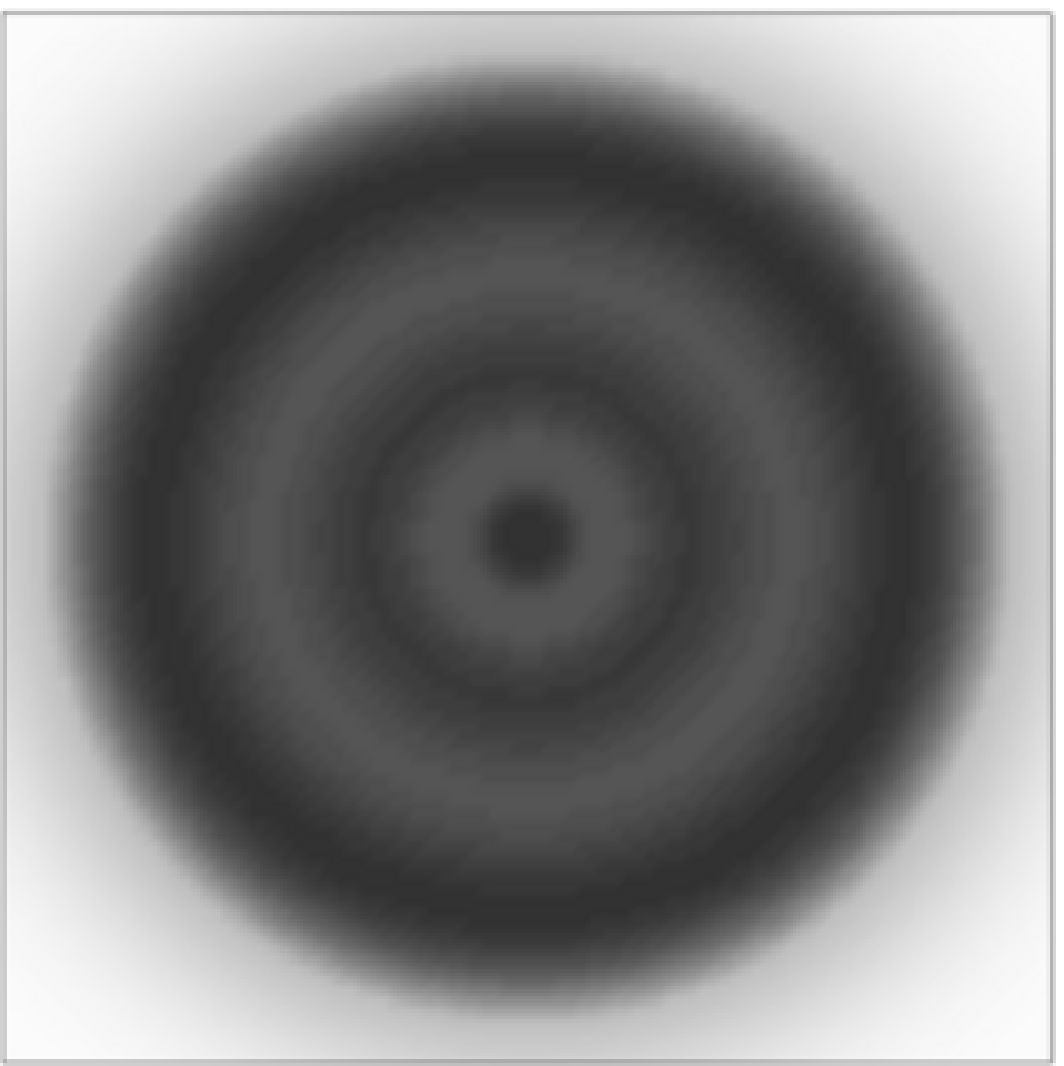}\\
\end{tabular}
%\end{center}
\caption{Classical  electron configuration (left) and quantum
mechanical electron density (right) for closed shell ground states
for N=2,6,12 20 and 30 electrons. The scale of all figures are from
-300 nm to 300 nm in both directions. \label{probabilities}}
\end{center}
\end{figure}

We now turn to a detailed comparison of the quantal ground state
probability distribution in absence of a magnetic field with
corresponding classical calculations. In Fig.~\ref{probabilities}
the classical electron configurations (left) with the
single-electron probability density obtained after integration over
the remaining $N-1$ electron's, $\rho(r_N) = \int dr_1 ... dr_{N-1}
|\Psi_N|^2$ (right) are shown. The classical frozen state is taken
directly from the final positions and placed on top of a figure with
each electron position rotating a number of uniformly distributed
angles with constant radius.

For the classical results of Fig.~\ref{probabilities},  the black
dots represent electron positions obtained from freezing an
initial distribution to $T'=0$. By keeping $T'>0$, but small, we
also produce a classical spatial density distribution by time
averaged superpositions of a large number of instantaneous
configurations as described in Ref.~\cite{hansen01}. These
densities are represented by shades of gray superimposed over the
static electron positions. The $N=6$ configuration is identical to
the one obtained in Ref.~\cite{classic93}. It contains a single
electron in the middle and five electrons in an outer ring. This
double structure is also seen for $N=12$, but now with 3 electrons
in the inner ring. The $N=19$ configuration not shown here but
published in \cite{classic93} contains the first triple ring with
one electron in the center, six electrons in the middle ring and
12 in the outer ring. It is interesting to note that when adding
one extra electron, the $N=20$ configuration minimizes energy by
placing the extra electron in the middle ring.

Comparison between classical and quantal probability density
distributions shows as expected a clear discrepancy for $N=2$
since anti symmetrization results in a spatial symmetric wave
function with a large probability of finding an electron in the
trap center. For $N=6$ the quantal ground state is a single ring
with  no inner electron, again due to antisymmetry, while the
classical $N=6$ configuration has a centrally positioned electron.
This pattern remains valid also for the remaining two cases, viz.,
where the classical calculations yield a  centrally placed
electron, the quantal calculations show a central density dip at
the  center, while the absence of central electron in the
classical configuration ($N=2$ and $N=12$) is matched by increased
central density in the quantal case. In the quantal displays the
gray shade coding has been selected to visualize the density
variations which are of the order of about $10-30\%$  in the
$N=12$ and $N=20$ cases. Apart from this effect of the Pauli
principle the systematics of the main shells  for $N = 6$, $N=12$
and $N=20$ compare well in spatial distribution with the
corresponding classical ones, showing that the classical static
configurations may roughly describe the ranges of the charge
density distributions even for systems with relatively few
confined electrons.

The grayscale in quantum densities is such that it reflects the
density variations which are increasingly smaller and smaller with
the number of electrons in the inner region (about 40 percent in the
N=6 case, about 20 for n=20 and around 10 percent for the largest
N-values. These variations reflect the nodal structure of the last
filled oscillator shell. The top or dip in the central region
follows the parity of the last filled shell which is negative for
N=6,20 and positive for N=12,30. For N=30 there are 3 visible
quantal and classical shells, the inner "inversion" is again
obvious. Half of the classical electrons occupy the outer shell
somewhat smaller than the quantal probability density for the reason
discussed above. We close by concluding that even if there are
distinct quantum features exposed in the right panels, the shell
structure and the sizes of the classical systems show a surprisingly
good correspondence with the quantal structures.

\section{Conclusion}

The present investigation has demonstrated the usefulness of the
Quantum Monte Carlo method for studies of relatively large numbers
electrons inside a quantum dot with parameters typically achieved in
experiments. Studies of the effects of the magnetic field have shown
that the energy per particle increase due to the magnetic field is
amplified by the electron repulsion, roughly proportional to the
number of electrons. his follows from comparisons with the
non-interacting cases and can be traced to the fact that the
repulsion leads to increased expectation values of the charge
distribution radius and the fact that the energetic effects of the
magnetic field depends on this quantity.

The quantal calculations were compared with classical Molecular
Dynamics calculations of the static equilibrium positions of
classical charges confined by a harmonic force. The comparisons for
closed shells shows that for the chosen type of quantum dots, the
effect of the Pauli principle leads to differing behavior of the
classical configurations and the quantal distributions in the
central area. The overall shell structure in the quantal and
classical case nevertheless has a remarkable agreement in shape and
spatial extension. This points further to the possibility that a
range of important dynamical processes in quantum dots may be
understood from purely classical models.

\begin{acknowledgments}
The present research has been partially sponsored by NFR through the
NANOMAT and the NOTUR programme.
\end{acknowledgments}

\end{document}